\documentclass{article}
\usepackage{amssymb}
\usepackage[T1]{fontenc}
\usepackage{balance}

\usepackage[preprint]{spconf}
\usepackage{amsmath,graphicx,hyperref}

\copyrightnotice{%
\parbox[b]{\textwidth}{\scriptsize
\copyright\ 2026 IEEE. Personal use of this material is permitted. Permission from IEEE must be obtained for all other uses, in any current or future media, including reprinting/republishing this material for advertising or promotional purposes, creating new collective works, for resale or redistribution to servers or lists, or reuse of any copyrighted component of this work in other works. Accepted for publication in Proc. IEEE ICASSP 2026. DOI: \href{https://doi.org/10.1109/ICASSP55912.2026.11462487}{10.1109/ICASSP55912.2026.11462487}.}%
}

\title{RIR-Former: Coordinate-Guided Transformer for Continuous Reconstruction of Room Impulse Responses}
\name{Shaoheng Xu$^{1*}$, Chunyi Sun$^{1*}$, Jihui (Aimee) Zhang$^{2,1}$, Prasanga Samarasinghe$^1$, Thushara Abhayapala$^1$ 
\thanks{* These authors contributed equally to this work.
\\ Code/data: \protect\url{https://github.com/ShaoHenry/RIR-Former}.
}}
\address{$^1$The Australian National University, Australia \\
$^2$The University of Queensland, Australia}
\begin{document}
\ninept
\maketitle
\begin{abstract}
Room impulse responses (RIRs) are essential for many acoustic signal processing tasks, yet measuring them densely across space is often impractical. In this work, we propose RIR-Former, a grid-free, one-step feed-forward model for RIR reconstruction. By introducing a sinusoidal encoding module into a transformer backbone, our method effectively incorporates microphone position information, enabling interpolation at arbitrary array locations. Furthermore, a segmented multi-branch decoder is designed to separately handle early reflections and late reverberation, improving reconstruction across the entire RIR. Experiments on diverse simulated acoustic environments demonstrate that RIR-Former consistently outperforms state-of-the-art baselines in terms of normalized mean square error (NMSE) and cosine distance (CD), under varying missing rates and array configurations. These results highlight the potential of our approach for practical deployment and motivate future work on scaling from randomly spaced linear arrays to complex array geometries, dynamic acoustic scenes, and real-world environments.
\end{abstract}


\vspace{-0.6cm}
\section{Introduction}
\label{sec:intro}
\vspace{-0.2cm}

Room Impulse Responses (RIRs) play a crucial role in acoustic signal processing. They encapsulate the acoustic characteristics of an environment and are essential for tasks such as:  
1) quantifying objective metrics for room design~\cite{RIR-ISO},  
2) enabling applications like sound source localization~\cite{RIR-SSL}, and  
3) supporting immersive experiences in virtual and augmented reality~\cite{AuralizationBook}.

However, measuring RIRs densely across a space is both time-consuming and labor-intensive. In complex environments, achieving spatially dense RIRs would require impractically extensive measurements. To address this, numerous RIR interpolation and reconstruction methods have been developed to estimate RIRs at unmeasured locations from a sparse set of measurements (e.g., GenDARA~\cite{GenDARA}). These approaches can be broadly categorized into traditional model-based and learning-based methods.

Traditional methods formulate RIR reconstruction via explicit mathematical models, including kernel ridge regression~\cite{Kernel,Kernel-2}, parametric sound field models~\cite{parametric-1,parametric-2}, and sparsity-driven models based on compressive sensing~\cite{cs-1,HSCMA,icasspCOMP,icasspSparsity,iwaencOMP,faHOA}. While effective under ideal conditions, these methods often struggle in acoustically challenging environments—such as those with long reverberation times.

Recent learning-based approaches leverage neural networks for RIR reconstruction~\cite{PIML-Review}. For example, Generative Adversarial Network (GAN)-based frameworks~\cite{GAN-sound} have been applied to reconstruct sound fields from spatially sparse, band-limited measurements. Convolutional Neural Networks (CNNs) have been used to reconstruct RIRs in uniform linear arrays (ULAs) by posing the problem as an inverse mapping~\cite{CNN-RIR}. However, such deep prior-based methods typically require retraining from scratch when the acoustic scene changes, limiting their practicality. To mitigate this,~\cite{dp-lora} incorporates Low-Rank Adaptation (LoRA) into a MultiResUNet framework, enabling efficient fine-tuning of pretrained CNN models across different scenarios. Separately,~\cite{PINN-NAH} proposes a Physics-Informed Neural Network (PINN) for near-field acoustic holography, embedding the wave equation as a constraint in the loss function. A similar PINN-based approach is adopted in~\cite{PINN-RIR} for continuous, grid-free RIR reconstruction without spatial discretization. Recently, Denoising Diffusion Probabilistic Models (DDPMs) have been applied to sound field reconstruction in the frequency domain~\cite{icassp-diffusion}, operating on individual frequency bins with experiments limited to 30–300~Hz—restricting applicability to full-band RIR reconstruction. 
Lastly, Implicit Neural Representation for Audio Scenes (INRAS)~\cite{INRAS} and Neural Acoustic Fields (NAFs)~\cite{LNAF} explore implicit neural representations of sound fields; the former requires explicit scene geometry as input, while the latter is typically learned via per-scene fitting.
Still, many existing methods face limitations: some focus only on low-frequency bands~\cite{icassp-diffusion}, others neglect phase information~\cite{GAN-sound}, or reconstruct only partial RIRs~\cite{icassp-diffusion,PINN-RIR}. Several approaches also rely on per-instance models that require retraining~\cite{CNN-RIR,PINN-RIR,LNAF} or adaptation~\cite{dp-lora} for each new acoustic environment, posing challenges for generalization.

More recently,~\cite{DiffusionRIR} reformulated RIR reconstruction as an image inpainting task by arranging microphone-by-time RIRs into grayscale images with missing rows, and applied a diffusion model for reconstruction. Although effective, this method is limited to uniformly spaced microphone arrays and fixed grid positions, making it unsuitable for arbitrary or irregular placements. The RIR is segmented along the time axis into square patches, which are mixed across different time segments during training. This disrupts the temporal structure, as early reflections and late reverberation follow different distributions. Such mixing weakens the distribution-matching objective of diffusion models. In addition, the UNet-based CNN lacks temporal awareness by treating all regions equally. The large number of denoising steps required during inference further limits practicality in real-time or low-latency scenarios.

In this work, we propose \textit{RIR-Former}, a transformer-based, one-step feed-forward RIR reconstruction method that is generalizable and computationally efficient. Unlike prior grid-based approaches, our model supports reconstruction at arbitrary positions along the array. To achieve this, we explicitly encode both the RIR measurements and their corresponding microphone positions as inputs to a transformer network. This allows the model to learn spatial dependencies and infer RIRs at unseen locations. In addition, we segment the RIR along the time axis into multiple pieces—capturing different stages of room reverberation—and use a multi-branch decoder to model these temporally distinct patterns more effectively.

We evaluate RIR-Former across diverse simulated acoustic environments with varying room sizes, reverberation times, microphone geometries, and source positions. The model demonstrates strong generalization and outperforms baselines in terms of Normalized Mean Squared Error (NMSE) and Cosine Distance (CD)~\cite{DiffusionRIR,CD}.

The key contributions of this paper are:  
1) a transformer-based model for \textbf{grid-free RIR reconstruction},  
2) a \textbf{segmented multi-branch architecture} tailored to different RIR components, and  
3) a \textbf{fast and generalizable feed-forward framework} suitable for practical deployment.

\begin{figure}[t]
  \centering
  \includegraphics[width=0.63\linewidth]{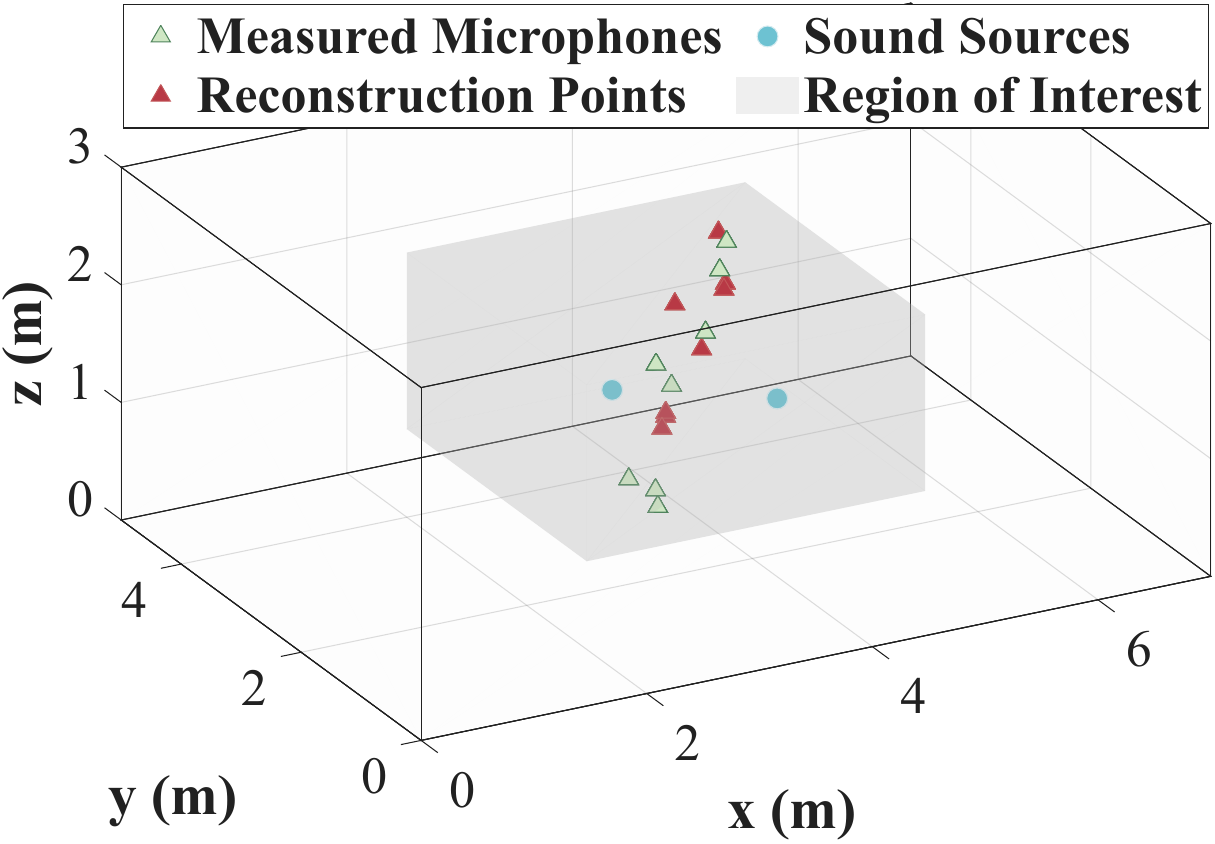}
  \vspace{-0.2cm}
  \caption{RIR reconstruction setup.}
  \label{fig:setup}
  \vspace{-0.53cm}
\end{figure}
\vspace{-0.2cm}
\section{Problem Formulation}
\label{sec:problem formulation}

The goal of this work is to reconstruct full RIRs at unmeasured locations based on a limited set of measured RIRs within a room.

Consider a general three-dimensional acoustic environment exhibiting reverberant characteristics. Let there be $M$ microphones located at positions $\mathbf{x}_m \equiv (x_m,\,y_m,\,z_m)$ for $m = 1, 2, \dots, M$, and $Q$ sources located at positions $\mathbf{y}_q \equiv (x_q,\,y_q,\,z_q)$ for $q = 1, 2, \dots, Q$. Additionally, let $N$ denote the number of unmeasured target positions, located at $\mathbf{z}_n \equiv (x_n,\,y_n,\,z_n)$ for $n = 1, 2, \dots, N$. All microphones, sources, and target positions are assumed to lie within a predefined 3D region of interest (ROI) denoted by $\Omega$. An illustration of this setup is shown in Fig.~\ref{fig:setup}.

At each microphone $\mathbf{x}_m$, the RIR is sampled at a frequency $f_s$ and truncated to $K$ samples, yielding a row vector $\mathbf{h}_m \in \mathbb{R}^{1 \times K}$. Stacking the $M$ measured RIRs forms the measurement matrix:
\begin{equation}
\label{eq:H}
\mathbf{H} = \begin{bmatrix} \mathbf{h}_1 & \mathbf{h}_2 & \cdots & \mathbf{h}_M \end{bmatrix}^\top \in \mathbb{R}^{M \times K},
\end{equation}
where $(\boldsymbol{\cdot})^\top$ denotes the transpose operation.

The positions of the microphones, sources, and target points can be stacked into:
\begin{align}
\mathbf{X}_m &= \begin{bmatrix} \mathbf{x}_1 & \mathbf{x}_2 & \cdots & \mathbf{x}_M \end{bmatrix}^\top \in \mathbb{R}^{M \times 3},\\
\mathbf{Y}_Q &= \begin{bmatrix} \mathbf{y}_1 & \mathbf{y}_2 & \cdots & \mathbf{y}_Q \end{bmatrix}^\top \in \mathbb{R}^{Q \times 3},\\
\mathbf{Z}_N &= \begin{bmatrix} \mathbf{z}_1 & \mathbf{z}_2 & \cdots & \mathbf{z}_N \end{bmatrix}^\top \in \mathbb{R}^{N \times 3}.
\end{align}

The objective is to estimate the RIRs at the $N$ unmeasured positions $\{\mathbf{z}_n\}$, with the same sampling rate $f_s$ and length $K$. Let $\mathbf{\bar{h}}_n$ and $\mathbf{\hat{h}}_n$ denote the ground truth and estimated RIRs at location $\mathbf{z}_n$, respectively. Concatenating these row vectors gives:
\begin{align}
\label{eq:Hbar}
\mathbf{\bar{H}} = \begin{bmatrix} \mathbf{\bar{h}}_1 & \mathbf{\bar{h}}_2 & \cdots & \mathbf{\bar{h}}_N \end{bmatrix}^\top \in \mathbb{R}^{N \times K}, \\
\label{eq:Hhat}
\mathbf{\hat{H}} = \begin{bmatrix} \mathbf{\hat{h}}_1 & \mathbf{\hat{h}}_2 & \cdots & \mathbf{\hat{h}}_N \end{bmatrix}^\top \in \mathbb{R}^{N \times K}.
\end{align}

The RIR reconstruction task can thus be formulated as follows:  
Given the measured RIR matrix $\mathbf{H}$ and the microphone positions $\mathbf{X}_m$, reconstruct the RIR matrix $\mathbf{\hat{H}}$ at unmeasured positions $\mathbf{Z}_N$ such that it closely approximates the ground truth $\mathbf{\bar{H}}$. This leads to the following optimization problem:
\begin{equation}
\label{eq:objective}
\mathbf{\hat{H}} = \arg\min_{\mathbf{\hat{H}}} \left\| \mathbf{\bar{H}} - \mathbf{\hat{H}} \right\|_2^2 \quad \text{given} \quad \mathbf{H},\, \mathbf{X}_m,\, \mathbf{Z}_N.
\end{equation}

In this work, we assume a shoebox room, a single fixed source, and all microphones and target positions lie coplanar within $\Omega$, interleaved along a linear array with either uniform (on-grid) or random (grid-free) spacing. Let $L = M + N$ denote the total number of array points. The missing rate (MR) is defined as $\text{MR} = N / L$.


\begin{figure*}[t!]
  \centering
  \includegraphics[width=0.9\linewidth]{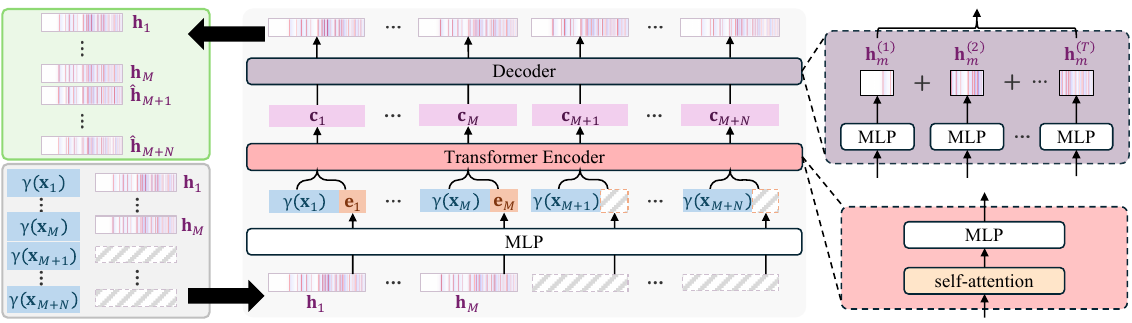}
  \vspace{-0.3cm}
  \caption{Known RIRs and their position embeddings are passed through an MLP for feature projection. The projected features are processed by a Transformer encoder, which captures spatial and contextual dependencies using self-attention mechanisms. The decoder, conditioned on the target position embedding \( x \) and learned feature map \( c \), consists of multiple MLPs that handle different segments of the RIR vector, followed by a final MLP that merges them to predict the unknown RIR at the desired location.}
  \label{fig:transformer}
\vspace{-0.4cm}
\end{figure*}

\vspace{-0.1cm}
\section{Proposed Method}
\label{sec:proposed method}
\vspace{-0.1cm}

Relying on handcrafted geometric priors is often inflexible; per-scene optimization is computationally expensive and lacks generalization; and treating the RIR as an image with local generative models imposes strong locality assumptions, emphasizing pattern completion over understanding spatial relationships. A more principled solution is to infer the room’s sound field from partial observations at known microphone positions and explicitly model its relationship with spatial location when querying unknown measurements. The Transformer architecture is well suited for this task, as it captures global dependencies and learns context-aware representations.

We propose RIR-Former, a Transformer-based model that learns spatial-acoustic dependencies from observed microphone RIRs and known geometry. As shown in Fig.~\ref{fig:transformer}, the architecture consists of a geometric encoder, a Transformer-based contextual reasoning module, a multi-channel segment-wise RIR decoder, and a residual refinement stage. This design enables scalable and generalizable RIR prediction across scenes with arbitrary sensor layouts and varying input sparsity, all within a single forward pass.

\noindent
\textbf{Geometric Encoding:}
For each microphone $\mathbf{x}_m$, it is passed through a sinusoidal positional encoding function to obtain positional token $\gamma(\mathbf{x}_m)$:
\begin{equation}
\begin{split}
\label{eq:encoding}
\gamma(\mathbf{x}_m) = [&\sin(2^0 \pi \cdot \mathbf{x}_m), \cos(2^0 \pi \cdot \mathbf{x}_m), \dots, \\
&\sin(2^{i-1} \pi \cdot \mathbf{x}_m), \cos(2^{i-1} \pi \cdot \mathbf{x}_m)],
\end{split}
\end{equation}
where $i = 6$. This encoding maps low-dimensional geometry into a richer space of periodic features, enabling the model to better capture both fine-scale and global spatial structures. Compared to using raw coordinates, this improves generalization to unseen geometries and scale variations.

\noindent
\textbf{Signal Encoding:}
Each observed RIR $\mathbf{h}_m \in \mathbb{R}^{1\times K}$ is projected into a latent feature vector $\mathbf{e}_m \in \mathbb{R}^{1\times D}$ using a learnable encoder, where $D$ is the chosen dimension of feature vector. When a measurement is missing, the input is zeroed and excluded from gradient updates. The final input token for each microphone is the concatenation $\mathbf{o}_m = [\gamma(\mathbf{x}_m); \mathbf{e}_m]$, combining spatial and acoustic information.

\noindent
\textbf{Transformer Encoder:}
The set of tokens $\{\mathbf{o}_m \}_{m=1}^M$ is passed through a multi-layer transformer encoder~\cite{nips-trans}. Through self-attention~\cite{nips-trans}, each microphone attends to all others, allowing the model to reason globally about how known responses inform missing ones. The output is a \textit{contextual microphone representation} $\mathbf{c}_m \in \mathbb{R}^{1\times D}$, which encodes geometry-aware, acoustically conditioned information at each microphone location. These representations capture complex patterns such as directional reflections, room-scale symmetry, and spatial redundancy across arbitrary microphone positions.

\noindent
\textbf{RIR Reconstruction:}
The contextual representation $\mathbf{c}_m$ is decoded into a full-length RIR $\hat{\mathbf{h}}_m \in \mathbb{R}^{1\times K}$ using a set of parallel MLP heads, each responsible for a different temporal segment of the RIR:
\begin{equation}
\label{eq:segment}
\hat{\mathbf{h}}_m = [\hat{\mathbf{h}}_m^{(1)} ; \hat{\mathbf{h}}_m^{(2)} ; \dots ; \hat{\mathbf{h}}_m^{(T)}].
\end{equation}
This segmented decoding enables temporal specialization across distinct acoustic regimes---e.g., direct sound, early reflections, and late reverberation---leading to improved reconstruction quality. A lightweight residual denoising module is applied to refine the output and reduce temporal artifacts.

\noindent
\textbf{Training Objective:}
During training, we randomly assign the $L$ array points as $M$ microphones and $N$ reconstruction targets based on the selected MR, resulting in $\mathbf{H}$ and $\mathbf{\bar{H}}$, respectively. The model is then optimized to reconstruct only the missing RIRs: $\mathbf{\hat{H}}$. The loss is defined as the mean squared error between the predictions and ground truth:
\begin{equation}
\mathcal{L} = \frac{1}{N} \| \hat{\mathbf{H}} - \mathbf{\bar{H}} \|_2^2,
\end{equation}
where $\hat{\mathbf{H}}$ is the model's prediction of $N$ RIRs.

Importantly, the model does not treat the RIRs as images. Each measured RIR, together with its microphone location, is converted into a unified token, allowing the Transformer to handle variable-length inputs and outputs corresponding to arbitrary numbers of known and unknown microphones. RIR-Former is trained with AdamW~\cite{adamW} using a learning rate of $3 \times 10^{-4}$, batch size 8, and for 200 epochs. During training, a special mechanism gradually increases the masking ratio from 30\% to 70\% over the first 10 epochs. We then finetune each individual decoder for prediction on each segment for 20 epochs; this effectively balances the imbalanced loss over the time dimension. We keep a high masking ratio during training, the model is encouraged to infer global contextual information rather than relying on local patterns. All RIRs are normalized per sample to ensure numerical stability.


\vspace{-0.1cm}
\section{Experiments}
\label{sec:experiments}
\vspace{-0.1cm}

In this section, we evaluate the RIR reconstruction performance of our proposed \textit{RIR-Former} through Monte Carlo simulations under diverse acoustic scenarios. We compare our method against three existing approaches.

\vspace{-0.3cm}
\subsection{Experiment Setup}
\label{ssec:experiment setup}
\vspace{-0.1cm}

We simulate realistic meeting room environments via Monte Carlo tests. A total of 8000 shoebox rooms are generated using~\cite{pyroomacoustics,ism,rir-g} to train the global models. Room length/width $\sim \mathcal{U}(4,8)$~m, height $\sim \mathcal{U}(2.5,4)$~m, and RT60 $\sim \mathcal{U}(0.2,0.8)$~s. The room center is set as the global origin $O$, and a planar ROI $\Omega$ is centered at $O$ with a single source inside. Along a linear array in $\Omega$, $L = 64$ points are placed and randomly assigned as $M$ microphones or $N$ targets to simulate different MR values. Each RIR is sampled at $f_s = 8000$~Hz.

\begin{figure}[t!]
  \centering
  \includegraphics[width=0.95\linewidth]{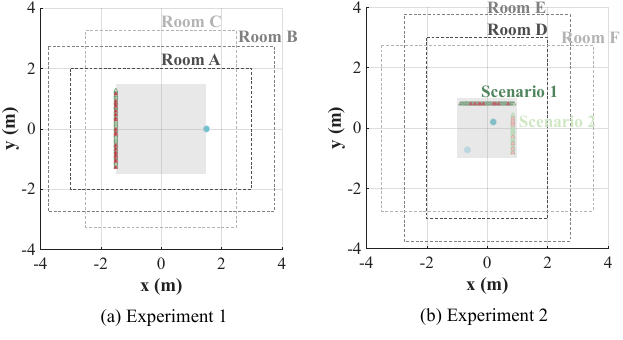}
  \vspace{-0.2cm}
  \caption{Experiment setups. (a) Experiment 1: fixed source, fixed array center, and uniform point spacing. (b) Experiment 2: random source position, random array placement, and randomized point spacing within the ROI $\Omega$.}
  \vspace{-0.42cm}
  \label{fig:setup_exp}
\end{figure}

We design two experiment setups with different levels of geometric randomness in array placement, point spacing, and source positions to evaluate robustness under increasing complexity:

\noindent
\textbf{Experiment 1 (Fixed Source, On-Grid, ULA):}
$M$ microphones and $N$ target points are uniformly spaced along a Uniform Linear Array (ULA) centered at $(-1.5,\,0,\,0)$. The array length is sampled uniformly between $1.28$~m and $3$~m. The source is fixed at $(1.5,\,0,\,0)$, and the ROI $\Omega$ is a $3\times3$~m square area (Fig.~\ref{fig:setup_exp}(a)). Each RIR is truncated to $K = 1024$ samples.

\noindent
\textbf{Experiment 2 (RSLA, Random Source, Grid-Free):}
To introduce higher variability, the array length, orientation, position, spacing, and source location are all randomized. The ROI $\Omega$ is a $2\times2$~m square. $M$ microphones and $N$ target points are placed along a Random-Spacing Linear Array (RSLA), oriented along either the $x$- or $y$-axis. The RSLA length is sampled uniformly between $1.28$~m and $2$~m, and its position within $\Omega$ is randomized. The source is also randomly located within $\Omega$ (Fig.~\ref{fig:setup_exp}(b)). Each RIR is truncated to $K = 2048$ samples.

\subsection{Comparison Methods}
\label{ssec:comparison method}

We compare our proposed RIR-Former against three existing methods: (1) \textbf{PINN}~\cite{PINN-RIR}, (2) \textbf{DiffusionRIR}~\cite{DiffusionRIR}, and (3) \textbf{Spline Cubic Interpolation (SCI)}~\cite{SCI}. To ensure a fair comparison, we replicated the DiffusionRIR approach using the OpenAI-enhanced diffusion network and confirmed that our implementation matched the performance reported in~\cite{DiffusionRIR} before retraining it on our dataset. For PINN, we directly used the authors' released code with our own data. For SCI, we implemented standard cubic spline interpolation along the spatial dimension.

\vspace{-0.2cm}
\subsection{Evaluation Metrics}
\label{ssec:evaluation metrics}

To quantify RIR reconstruction quality, we use two metrics:  
(1) \textbf{Normalized Mean Squared Error (NMSE)} and  
(2) \textbf{Cosine Distance (CD)}~\cite{DiffusionRIR,CD}. The metrics are defined as:
\begin{equation}
\text{NMSE} = 10\log_{10}\!\left(
\tfrac{\|\bar{\mathbf{H}}-\hat{\mathbf{H}}\|_F^2}{\|\bar{\mathbf{H}}\|_F^2}
\right),~
\text{CD} = \tfrac{1}{N}\!\sum_{n=1}^N\!\left(
1-\tfrac{\bar{\mathbf{h}}_n \hat{\mathbf{h}}_n^T}{\|\bar{\mathbf{h}}_n\|_2 \|\hat{\mathbf{h}}_n\|_2}
\right),
\end{equation}
where $\|\boldsymbol{\cdot}\|_F$ denotes the Frobenius norm and $\|\boldsymbol{\cdot}\|_2$ denotes the Euclidean norm. Lower values of NMSE and CD indicate better reconstruction quality. CD measures the scale-invariant waveform similarity (i.e., shape alignment) between the predicted and ground truth RIRs and ranges from $0$ to $2$, where CD = 0 implies identical waveform shapes. We also evaluate the average retraining and inference time of each method on a new acoustic scene using an NVIDIA A100 GPU, and report the results in Table~\ref{tab:exp1-res}.

\vspace{-0.2cm}
\subsection{Results}
\label{ssec:results}

\begin{figure}[t]
  \centering
  \includegraphics[width=1\linewidth]{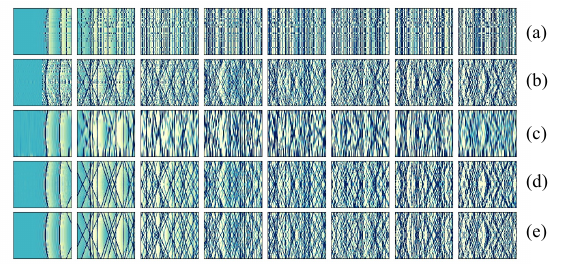}
  \caption{RIR reconstruction results, segmented into 8 parts and normalized within each segment for better visualization. The subfigures compare different methods: (a) SCI, (b) DiffusionRIR, (c) PINN, (d) Ours, and (e) Ground Truth.}
  \label{fig:cmp_result}
\end{figure}

\begin{figure}[t]
  \centering
  \includegraphics[width=1\linewidth]{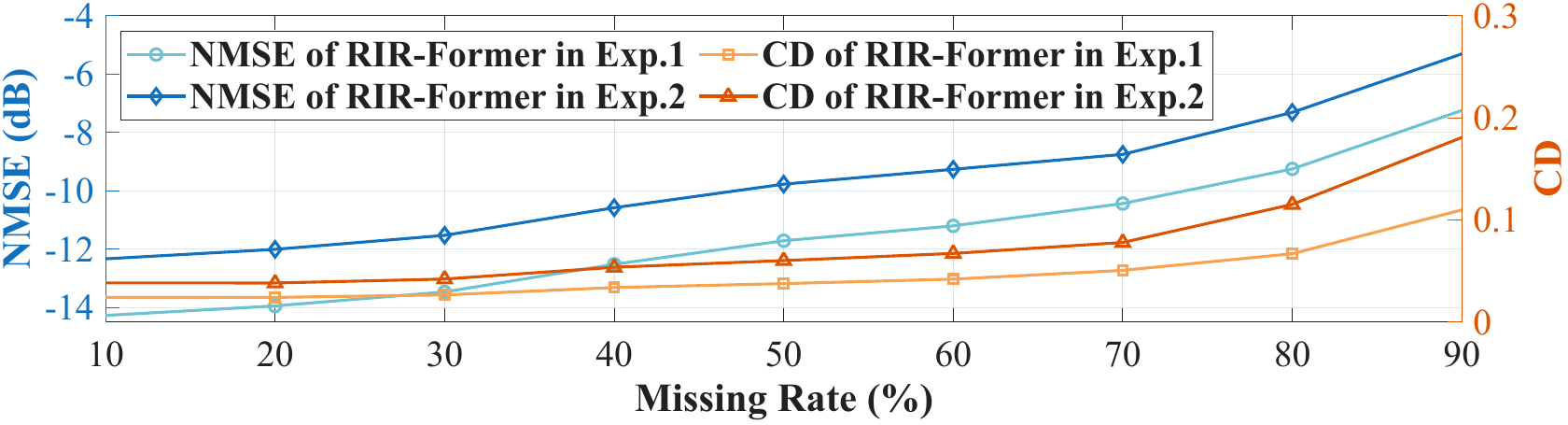}
  \vspace{-0.4cm}
  \caption{NMSE and CD across different missing rates (Exp.~1~and~2).}
  \label{fig:NMSE_CD}
\end{figure}
In both Experiment 1 and 2, we evaluate the NMSE and CD of RIR-Former across 10 simulated acoustic environments, each tested under missing rates from $10\%$ to $90\%$.  For each rate, we average the results across all environments. As shown in Fig.~\ref{fig:NMSE_CD}, both NMSE and CD slightly degrade with increasing missing rates but remain consistently low. Notably, the NMSE stays below $-5$ dB and CD remains below $0.2$ even at a $90\%$ missing rate, demonstrating the robustness and generalization capability of our method.

In Experiment 1 (on-grid ULA), We further compare RIR-Former with three baselines under a fixed missing rate of $70\%$. Table~\ref{tab:exp1-res} summarizes the results. RIR-Former achieves the lowest NMSE and CD among all methods. Fig.~\ref{fig:cmp_result} presents a visual comparison of the reconstructed RIRs, segmented into 8 temporal regions for clarity. Our method (d) consistently matches the ground truth (e) across all segments, while DiffusionRIR (b) reconstructs the late reverberation well but introduces noise. PINN (c) achieves moderate performance but suffers from visible artifacts. SCI (a) fails to model the RIR structure after the initial two segments. Furthermore, RIR-Former does not require retraining for new acoustic scenes and achieves the fastest inference time, owing to its one-step feed-forward architecture.

In the more complex Experiment 2 (grid-free RSLA), only PINN and our method are compared since DiffusionRIR and SCI are restricted to fixed-grid reconstruction. Table~\ref{tab:exp2-res} shows that RIR-Former continues to outperform PINN significantly under a $70\%$ missing rate. Interestingly, PINN shows a slight performance improvement from Experiment 1 to 2. This is likely because PINN, as a data-fitting method, is less sensitive to scene complexity.

Lastly, we conduct two ablation studies based on Experiment 2:  
(1) removing the sinusoidal encoding module from Eq.~\eqref{eq:encoding}, and  
(2) removing the segmented multi-branch decoder from Eq.~\eqref{eq:segment}.  
Table~\ref{tab:ablation} presents the results. The removal of sinusoidal encoding increases NMSE by 3.97 dB, while removing the segment decoder leads to a 2.24 dB degradation. This confirms the effectiveness of both design components. The sinusoidal encoding enriches the geometric input representation, while the segment-wise decoder balances optimization between early and late RIR components, avoiding bias toward high-energy early reflections.

\vspace{-0.4cm}
\begin{table}[h]
  \centering
  \caption{Quantitative results for Experiment 1 (ULA, fixed source)}
  \label{tab:exp1-res}
  \begin{tabular}{c|c|c|c|c}
    \hline
    Method        & NMSE (dB) & CD     & Re-train            & Inference\\
    \hline
    Ours          & -10.440   & 0.051  & N/A                 &  0.002~s\\
    PINN          & -2.557    & 0.293  & $\geq1~\text{hour}$ &  0.883~s\\
    DiffusionRIR  & -0.618    & 0.325  & N/A                 &  128.8~s\\
    SCI           & 2.170     & 0.808  & N/A                 &  0.178~s\\
    \hline
  \end{tabular}
\vspace{-0.8cm}
\end{table}

\begin{table}[h]
  \centering
  \caption{Results for Experiment 2 (RSLA, random source).}
  \label{tab:exp2-res}
  \begin{tabular}{c|c|c}
    \hline
    Method                & NMSE (dB) & CD \\
    \hline  
    Ours                  & -8.755 & 0.078 \\
    PINN                  & -3.158 & 0.319 \\
    \hline
  \end{tabular}
\vspace{-0.8cm}
\end{table}

\begin{table}[h]
  \centering
  \caption{Ablation study results based on Experiment 2.}
  \label{tab:ablation}
  \begin{tabular}{c|c|c}
    \hline
    Method                & NMSE (dB) & CD \\
    \hline
    w/o. sinusoidal encoding&-4.781&0.177     \\
    w/o. segment decoder&-6.516&0.118 \\ 
    \hline
  \end{tabular}
\vspace{-0.6cm}
\end{table}


\section{Conclusion}
\label{sec:conclusion}
\vspace{-0.2cm}

In this paper, we proposed a grid-free, one-step feed-forward model for RIR reconstruction. By incorporating a sinusoidal encoding module into a Transformer architecture, our model effectively encodes microphone positions, enabling accurate reconstruction at arbitrary spatial locations. The segmented multi-branch decoder balances the importance of early and late reflections, yielding high-quality reconstruction across the entire RIR. Ablation studies validate the effectiveness of each component. Through extensive simulations, our method demonstrates superior performance in terms of NMSE and CD, outperforming three state-of-the-art baselines under various acoustic conditions. Future work includes extending the system to complex microphone array geometries, dynamic acoustic scenes, and validating its performance on real-world data.


\vfill\pagebreak

\balance
\bibliographystyle{IEEEbib}
\bibliography{strings,refs}

\end{document}